\documentclass{tlp}

\usepackage{epsfig}
\usepackage{xspace}
\usepackage{wrapfig}
\usepackage{listings}
\usepackage{latexsym,amssymb}
\usepackage{amsmath}
\usepackage{alltt}
\usepackage[all]{xy}
\usepackage{multirow}
\usepackage{color} 
\usepackage{graphics}
\usepackage{graphicx}
\usepackage[draft]{fixme}
\usepackage{booktabs}
\usepackage{bold-extra}

\usepackage{fancyhdr} 

\def\tuple#1{\langle #1 \rangle}

\newcommand{\gd}[0]{~\rule{0.5mm}{2.5mm}~}

\newcommand{\clp}{CLP\xspace}

\newcommand{\dpk}[1]{{\bf depth-$#1$}\xspace}
\newcommand{\bck}[1]{{\bf block-$#1$}\xspace}

\newcommand{\cc}{CC\xspace}
\newcommand{\ccs}{CCs\xspace}
\newcommand{\newobject}{\textsf{new\_object}}
\newcommand{\newarray}{\textsf{new\_array}}
\newcommand{\getfield}{\textsf{get\_field}}
\newcommand{\setfield}{\textsf{set\_field}}
\newcommand{\getarray}{\textsf{get\_array}}
\newcommand{\setarray}{\textsf{set\_array}}
\newcommand{\type}{\textsf{type}}
\newcommand{\length}{\textsf{length}}
\newcommand{\subclass}{\textsf{subclass}}
\newcommand{\getcell}{\textsf{get\_cell}}
\newcommand{\setcell}{\textsf{set\_cell}}
\newcommand{\memberdet}{\textsf{member\_det}}
\newcommand{\replacedet}{\textsf{replace\_det}}

\newcommand{\secbeg}{\vspace*{-0cm}}

\newcommand{\tdg}{TDG\xspace} 
\newcommand{\tcg}{TCG\xspace} 
\newcommand{\syex}{SymEx\xspace} 
\newcommand{\pet}{PET\xspace}

\def\anno#1{{\ooalign{\hfil\raise.07ex\hbox{\small{\rm \textcolor{red}{#1}}}\hfil%
        \crcr{\small \textcolor{blue}{\mathhexbox20D}}}}}
\def\sanno#1{{\ooalign{\hfil\raise.07ex\hbox{\tiny{\rm \textcolor{red}{#1}}}\hfil%
        \crcr{\tiny \textcolor{blue}{\mathhexbox20D}}}}}

\newtheorem{definition}{Definition} 
\newtheorem{example}{Example} 
\newtheorem{theorem}{Theorem} 

\begin{document}

\title{Test Case Generation for Object-Oriented Imperative Languages
  in CLP}

\author[M. G\'omez-Zamalloa, E. Albert and G. Puebla]
{MIGUEL G\'OMEZ-ZAMALLOA$^{1}$, ELVIRA ALBERT$^{1}$, and
   GERM\'AN PUEBLA$^{2}$ \\ \\
$^{1}$ DSIC, Complutense University of Madrid (UCM), E-28040 Madrid, Spain\\
$^{2}$ DLSIIS, Technical University of Madrid (UPM), E-28660 Boadilla del
Monte, Madrid, Spain\\
}

\setcounter{page}{1}

\maketitle



\vspace{-.3cm}
\begin{abstract}
  Testing is a vital part of the software development process. Test
  Case Generation (\tcg) is the process of automatically generating a
  collection of test-cases which are applied to a system under test.
  White-box \tcg is usually performed by means of \emph{symbolic
    execution}, i.e., instead of executing the program on normal
  values (e.g., numbers), the program is executed on symbolic values
  representing arbitrary values.  When dealing with an object-oriented
  (OO) imperative language, symbolic execution becomes challenging as,
  among other things, it must be able to backtrack, complex
  heap-allocated data structures should be created during the \tcg
  process and features like inheritance, virtual invocations and
  exceptions have to be taken into account. Due to its inherent
  symbolic execution mechanism, we pursue in this paper that
  \emph{Constraint Logic Programming} (CLP) has a promising
   application field in \tcg.  We will support our claim by
  developing a fully CLP-based framework to \tcg of an OO imperative
  language, and by assessing it on a corresponding implementation on a
  set of challenging Java programs.  
\end{abstract} 
\vspace{-.2cm}


\secbeg

\section{Introduction}
\label{sec:introduction}


Test Case Generation (\tcg) is the process of automatically generating
a collection of test-cases which are applied to a system under test.
The generated cases must ensure a certain \emph{coverage criterion}
(see e.g.,~\cite{267590} for a survey) which are heuristics that
estimates how well the program is exercised by a test suite. Examples
of coverage criteria are \emph{statement coverage}, which requires
that each line of the code is executed, \emph{path coverage} which
requires that every possible trace through a given part of the code is
executed, \emph{loop-$k$} (resp. \emph{block-$k$}) which limits to a
threshold $k$ the number of times we iterate on loops (resp. visit
blocks in the control flow graph~\cite{tdg-lopstr08}).
Among all possible forms of \tcg, we focus on \emph{static} (i.e., no
knowledge about the input data is assumed) and \emph{white-box} \tcg
(i.e., the program is used for guiding the \tcg process).
The standard way of performing static white-box \tcg is program
\emph{symbolic execution}
(\syex)~\cite{360252,GotliebBR00-short,Meudec01,MullerLK04,DBLP:conf/tap/TillmannH08},
whereby instead of on actual values, programs are executed on symbolic
values, sometimes represented as \emph{constraint variables}.
Such constraints are accumulated into \emph{path constraints} as each
path of the execution tree is expanded.  The path constraints in
feasible paths provide pre-conditions on the input data which
guarantee that the corresponding path will be executed at run-time.
%
%

In this paper, we pursue that \emph{Constraint Logic Programming}
(CLP) has a promising application field in \tcg, since
it inherently combines the use of constraint solvers into its \syex
mechanism. Our main goal is to formalize a whole \tcg framework for a
realistic object-oriented (OO) imperative language by means of
CLP. Our approach consists of two basic parts: first, the imperative
program is compiled into an equivalent CLP program and, second, \tcg
is performed on the CLP program by relying only on CLP's evaluation
mechanisms.
The main challenges in \tcg when dealing with an OO imperative
language are related to the creation of complex heap-allocated data
structures during the \tcg process, and to handling OO features like
inheritance and virtual invocations, and exceptions.
%
Besides, when dealing with objects, one needs to take into account all
possible \emph{aliasing} among them, since this might affect directly
the coverage of the test-cases.  Previous approaches strive to define
novel specific constraint operators to carry out these tasks (see e.g.
\cite{CharreteurBG09,Schrijvers-LOPSTR09}).  Instead, in our approach,
the whole \tcg process is formulated using CLP only, and without the
need of defining specific operators to handle the different
features. This, on one hand, has the advantage of providing a clean
and uniform formalization. And, more importantly, since \syex is
performed on an equivalent CLP program, we can often obtain the
desired degree of coverage by using existing evaluation strategies on
the CLP side.  This gives us flexibility and
parametricity w.r.t.\ the adequacy criteria.
   
Our approach has been integrated in \pet \cite{tdg-pepm10}, a
Partial-Evaluation based \tcg tool, 
extending its applicability towards real-life OO applications.

\secbeg

\secbeg

\section{A CLP-Executable Object-Oriented Imperative Language}\label{sec:clp-exec-object}

\begin{figure}[t]
\begin{center}
\setlength\fboxsep{2pt}
\footnotesize
\fbox{
\begin{tabular*}{0.96\textwidth}{@{\extracolsep{\fill}}@{\!\!}l|@{\!\!}l}
\( \!\!
\begin{array}{l}
\mbox{\tt class SLNode \{}\\
\mbox{\tt ~~~int data;}\\
\mbox{\tt ~~~SLNode next;}\\
\mbox{\tt \}}\\ \\
\mbox{\tt class SortedList \{}\\
\mbox{\tt ~~~SLNode first;}\\ \\
\mbox{\tt ~~~public void merge(SortedList l)\{}\\
\mbox{\tt ~~~~~~SLNode p1,p2,curr;}\\
\mbox{\tt ~~~~~~p1 = first; p2 = \anno{1} l.first;}\\
\mbox{\tt ~~~~~~if (\anno{2} p1.data <= \anno{3} p2.data)} \emph{// if1} \\ 
\mbox{\tt ~~~~~~~~~p1 = \anno{4}  p1.next;}\\
\mbox{\tt ~~~~~~else \{}\\
\mbox{\tt ~~~~~~~~~first = p2; p2 = p2.next; \}}\\
\mbox{\tt ~~~~~~curr = first;} \emph{~~~// preloop}\\
\end{array}
\)
&
\(
\begin{array}{l}
\mbox{\it // loop}\\
\mbox{\tt ~~while ((p1 != null) \&\& (p2 != null))\{}\\
\mbox{\it ~~~// loopcond1, loopcond2 and loopbody1}\\
\mbox{\tt ~~~~~if (p1.data <= p2.data)\{}   \emph{// if2} \\
\mbox{\tt ~~~~~~~~curr.next = p1;}\\
\mbox{\tt ~~~~~~~~p1 = p1.next;}\\
\mbox{\tt ~~~~~\}}\\
\mbox{\tt ~~~~~else \{}\\
\mbox{\tt ~~~~~~~~curr.next = p2;}\\
\mbox{\tt ~~~~~~~~p2 = p2.next;}\\
\mbox{\tt ~~~~~\}}\\
\mbox{\tt ~~~~~curr = curr.next;} \emph{// loopbody2}\\
\mbox{\tt ~~\}}\\
\mbox{\tt ~~if (p1 == null) curr.next = p2;} \emph{// if3} \\
\mbox{\tt ~~else curr.next = p1;}\\
\mbox{\tt \}}\\
\end{array}
\)

\end{tabular*}
}
\end{center}
\secbeg\secbeg
\caption{Working example: Java source code}
\label{fig:example-java}
\end{figure}

In this section, we define the (CLP) syntax and semantics of the OO
imperative language on which our \tcg approach is developed, which we
call \emph{CLP-decompiled} language.
Its main characteristic is that it keeps all features of the original
OO language but it is \emph{CLP-executable}, i.e., it can be executed
using the evaluation mechanism of CLP languages. When the source
imperative language is low-level as bytecode, we use the term
CLP-\emph{de}compiled language. In previous work, it has been shown
that Java bytecode (and hence Java) can be decompiled into a similar
language \cite{mod-decomp-jist09} by relying on the interpretive
approach \cite{Futamura:71:54} to compilation, proposed in the first
Futamura projection. In this approach, the CLP-(de)compilation is
obtained by partially evaluating an interpreter for the OO language
written in CLP. 

\begin{example}
  Fig.~\ref{fig:example-java} shows the source code of our running
  example which implements a merge algorithm on sorted singly-linked
  lists. Fig.~\ref{fig:example-clp} shows the CLP-decompiled program
  automatically generated by our system from the bytecode obtained by
  compiling the Java program, with some simplifications to improve
  readability. The correspondence between blocks of the original
  program and clauses in the decompiled one is shown in comments in
  the Java code.  The main features that can be observed from the
  decompilation are: (1) All clauses contain input and output
  arguments and heaps, and an exceptional flag. As in the bytecode,
  input arguments of non-static methods include the reference
  \emph{this} (named $\sf r(Th)$). Reference variables are of the form
  $\sf r(V)$ and we use the same variable name $\sf V$ as in the
  program.  
  (2) Java exceptions are made explicit in the decompiled
  program. Observe predicates $\tt nullcheck$$x$, which capture the
  exceptions that can be thrown at program points annotated as
  \anno{x}.
%
  (3) Conditional statements in the source program are transformed to
  guarded rules in the CLP one (e.g., $\sf if1$). (4) Iteration in the
  source program is transformed into recursion in the CLP
  program. E.g, the while loop corresponds to the recursive predicate
  $\sf loop$.
\end{example}

\begin{figure}[t]
\begin{center}
\setlength\fboxsep{2pt}
\footnotesize
\fbox{
\begin{tabular*}{0.975\textwidth}{@{\extracolsep{\fill}}l}
\( \hspace{-.3cm}
\begin{array}{l}
\mbox{\tt merge([[r(Th),L],[],H$_{in}$,H$_{out}$,EF) :- 
   \textbf{\getfield}(Hin,Th,'SL':first,P1),}\\
\mbox{\tt ~~~~~~~~nullcheck1([r(Th),L,P1],[],H$_{in}$,H$_{out}$,EF).}\\
\\
\mbox{\tt nullcheck1$_1$([r(Th),r(L),P1],[],H1,H2,EF) :- 
   \textbf{\getfield}(H1,L,'SL':first,P2),}\\
\mbox{\tt ~~~~~~~~nullcheck2([r(Th),r(L),P1,P2],[],H1,H2,EF).}\\
\mbox{\tt nullcheck1$_2$([r(\_),null,\_],[],H1,H2,exc(ExRef)) :- 
   \textbf{\newobject}(H1,'NPE',ExRef,H2).}\\
\\
\mbox{\tt nullcheck2$_1$([r(Th),r(L),r(P1),P2],[],H1,H2,EF) :- 
   \textbf{\getfield}(H1,P1,'SL':data,Data1),}\\
\mbox{\tt ~~~~~~~~nullcheck3([Data1,r(Th),r(L),r(P1),P2],[],H1,H2,EF).}\\
\mbox{\tt nullcheck2$_2$([r(\_),r(\_),null,\_],[],H1,H2,exc(ExRef)) :- 
   \textbf{\newobject}(H1,'NPE',ExRef,H2).}\\
\\
\mbox{\tt nullcheck3$_1$([D1,r(Th),r(L),r(P1),r(P2)],[],H1,H2,EF) :- 
   \textbf{\getfield}(H1,P2,'SL':data,D2),}\\
\mbox{\tt ~~~~~~~~if1([D2,D1,r(Th),r(L),r(P1),r(P2)],[],H1,H2,EF).}\\
\mbox{\tt nullcheck3$_2$([\_,\_,r(\_),r(\_),null],[],H1,H2,exc(ExR))
  :- \textbf{\newobject}(H1,'NPE',ExR,H2).}\\
\\
\mbox{\tt if1$_1$([Data2,Data1,r(Th),r(L),r(P1),r(P2)],[],H1,H3,EF) :- Data1 \#> Data2,}\\
\mbox{\tt ~~~~~~~~\textbf{\setfield}(H1,Th,'SL':first,r(P2),H2), 
                  \textbf{\getfield}(H2,P2,'SL':next,P2'),}\\
\mbox{\tt ~~~~~~~~preloop([r(Th),r(L),r(P1),P2'],[],H2,H3,EF).}\\
\mbox{\tt if1$_1$([Data2,Data1,r(Th),r(L),r(P1),r(P2)],[],H1,H2,EF) :- Data1 \#=< Data2,}\\
\mbox{\tt ~~~~~~~~\textbf{\getfield}(H1,P1,'SL':next,P1'), 
                  preloop([r(Th),r(L),P1',r(P2)],[],H1,H2,EF).}\\
\\
\mbox{\tt preloop([r(Th),L,P1,P2],[],H1,H2,EF) :-}\\
\mbox{\tt ~~~~~~~~\textbf{\getfield}(H1,Th,'SL':first,Curr),
   loop([r(Th),L,P1,P2,Curr],[],H1,H2,EF).}\\
\\
\mbox{\tt loop([Th,L,P1,P2,Curr],[],H1,H2,EF) :- 
   loopcond1([Th,L,P1,P2,Curr],[],H1,H2,EF).}\\

\mbox{\tt loopcond1$_1$([\_,\_,null,P2,Curr],[],H1,H2,EF) :- 
   if3([null,P2,Curr],[],H1,H2,EF).}\\
\mbox{\tt loopcond1$_2$([Th,L,r(P1),P2,Curr],[],H1,H2,EF) :-}\\
\mbox{\tt ~~~~~~~~loopcond2([Th,L,r(P1),P2,Curr],[],H1,H2,EF).}\\
\\
\mbox{\tt loopcond2$_1$([\_,\_,r(P1),null,Cur],[],H1,H2,EF) :- 
   if3([r(P1),null,Cur],[],H1,H2,EF).}\\
\mbox{\tt loopcond2$_2$([Th,L,r(P1),r(P2),Curr],[],H1,H2,EF) :-}\\
\mbox{\tt ~~~~~~~~loopbody1([Th,L,r(P1),r(P2),Curr],[],H1,H2,EF).}\\
\\
\mbox{\tt loopbody1([Th,L,r(P1),r(P2),Curr],[],H1,H2,EF) :-}\\
\mbox{\tt ~~~~~~~~\textbf{\getfield}(H1,P1,'SL':data,Data1), \textbf{\getfield}(H1,P2,'SL':data,Data2),}\\
\mbox{\tt ~~~~~~~~if2([Data2,Data1,Th,L,r(P1),r(P2),Curr],[],H1,H2,EF).}\\
\\
\mbox{\tt if2$_1$([Data2,Data1,Th,L,r(P1),r(P2),r(Curr)],[],H1,H3,EF) :- Data1 \#> Data2,}\\
\mbox{\tt ~~~~~~~~\textbf{\setfield}(H1,Curr,'SL':next,r(P2),H2), 
                  \textbf{\getfield}(H2,P2,'SL':next,P2'),}\\
\mbox{\tt ~~~~~~~~loopbody2([Th,L,r(P1),P2',r(Curr)],[],H2,H3,EF).}\\
\mbox{\tt if2$_2$([Data2,Data1,Th,L,r(P1),r(P2),r(Curr)],[],H1,H3,EF) :- Data1 \#=< Data2,}\\
\mbox{\tt ~~~~~~~~\textbf{\setfield}(H1,Curr,'SL':next,r(P1),H2), 
                  \textbf{\getfield}(H2,P1,'SL':next,P1'),}\\
\mbox{\tt ~~~~~~~~loopbody2([Th,L,P1',r(P2),r(Curr)],[],H2,H3,EF).}\\
\\
\mbox{\tt loopbody2([Th,L,P1,P2,r(Curr)],[],H1,H2,EF) :-}\\
\mbox{\tt ~~~~~~~~\textbf{\getfield}(H1,Curr,'SL':next,Curr'), 
                  loop([Th,L,P1,P2,Curr'],[],H1,H2,EF).}\\
\\
\mbox{\tt if3$_1$([r(P1),\_,r(Curr)],[],H1,H2,ok) :- 
   \textbf{\setfield}(H1,Curr,'SL':next,r(P1),H2).}\\
\mbox{\tt if3$_2$([null,P2,r(Curr)],[],H1,H2,ok) :- 
   \textbf{\setfield}(H1,Curr,'SL':next,P2,H2).}\\

\end{array}
\)
\end{tabular*}
}
\end{center}
\secbeg\secbeg
\caption{Working example: \clp-decompiled code}
\label{fig:example-clp}
\end{figure}

\secbeg
\subsection{Syntax of CLP-Decompiled Object-Oriented Imperative
  Programs}\label{sec:synt-clp-decomp}

As illustrated in Fig.~\ref{fig:example-clp}, a \emph{CLP-decompiled
  program} consists of a set of \emph{predicates}.
A predicate $p$ is defined by one or more clauses which are mutually
exclusive. This is ensured, either by means of mutually exclusive
\emph{guards}, or by information made explicit on the clause heads (as
usual in \clp). Each clause $p$ receives as input a (possibly empty)
list of arguments \emph{Args$_{in}$} and an input heap
\emph{H$_{in}$}, and returns the (possibly empty) output
\emph{Args$_{out}$}, a possibly modified output heap \emph{H$_{out}$},
and an exception flag. This flag indicates whether the execution ends
normally or with an uncaught exception. Clauses adhere to the
following grammar. As usual, terminals start by lowercase (or special
symbols) and non-terminals by uppercase. Subscripts are provided just
for clarity.
\[ \small \hspace{-.35cm}
\begin{array}{l}
\begin{array}{r@{\,}l@{\,}l}
Clause & ::= & \mbox{{\it Pred} ({\it Args$_{in}$,Args$_{out}$,H$_{in}$,H$_{out}$,ExFlag}) 
                    :- {\it [G,]B$_1$,B$_2$,\ldots,B$_n$.}} \\
G & ::= & \mbox{{\it Num* ROp Num*} $|$ {\it Ref$_1^*$} \textbackslash == {\it Ref$_2^*$} } 
    \mbox{$|$ \type({\it H,Ref$^*$,T})} \\ 

B & ::= & \mbox{{\it Var} \#= {\it Num* AOp Num*} $|$
              {\it Pred} ({\it
                Args$_{in}$,Args$_{out}$,H$_{in}$,H$_{out}$,ExFlag}) $|$}\\
  & & \mbox{\newobject({\it H,C$^*$,Ref$^*$,H}) $|$
      \newarray({\it H,T,Num$^*$,Ref$^*$,H}) $|$ \length({\it H,Ref$^*$,Var}) $|$}\\
  & & \mbox{\getfield({\it H,Ref$^*$,FSig,Var}) $|$ 
            \setfield({\it H,Ref$^*$,FSig,Data$^*$,H}) $|$}\\
  & & \mbox{\getarray({\it H,Ref$^*$,Num$^*$,Var}) $|$ 
            \setarray({\it H,Ref$^*$,Num$^*$,Data$^*$,H})}\secbeg\secbeg\\
\end{array} \\ \\
\end{array}
\]

\[ \small \hspace{-.35cm}
~~~\begin{array}{ll}
\begin{array}{r@{\,}l@{\,}l}
Pred & ::= & \mbox{{\it Block} $|$ {\it MSig}} \\
Args & ::= & \mbox{[] $|$ [{\it Data$^*|$Args}]} \\
Data & ::= & \mbox{\it Num $|$ Ref} \\
Ref & ::= & \mbox{\it null $|$ r(Var)} \\
\mbox{\it ExFlag} & ::= & \mbox{\it ok $|$ exc(Var)} \\
\end{array}
 ~ & ~
\begin{array}{r@{\,}l@{\,}l}
ROp & ::= & \mbox{ \#$>$ $|$ \#$<$ $|$ \#$>=$ $|$ \#$=<$  $|$ \#$=$
                       $|$ \#\textbackslash = } \\
AOp & ::= & \mbox{ + $|$ - $|$ $*$ $|$ / $|$ {\it mod} } \\
T & ::= & \mbox{\it bool $|$ int $|$ C $|$ array(T)} \\ 
\mbox{\it FSig} & ::= & \mbox{\it C$:$FN} \\ 
H & ::= & \mbox{\it Var} \\
\end{array}
\end{array}
\]
Non-terminals {\it Block}, {\it Num}, {\it Var}, {\it FN}, {\it MSig}
and {\it C} denote, resp., the set of predicate names, numbers,
variables, field names, method signatures and class names. Observe
that clauses can define both methods which appear in the original
source program ({\it MSig}), or additional predicates which correspond
to intermediate blocks in the program ({\it Block}).  An asterisk on a
non-terminal denotes that it can be either as defined by the grammar
or a (possibly constraint) variable.
Guards might contain: comparisons between numeric data or references
and calls to the \type\ predicate, which checks the type of a
reference variable (by consulting the heap).  Virtual method
invocations in the OO language are resolved at compile-time by looking
up all possible runtime instances of the method. In the decompiled
program, they are translated into a choice of \type\ instructions
which check the actual object type, followed by the corresponding
method invocation for each runtime instance.
Instructions in the body of clauses include: (first row) arithmetic
operations, calls to other predicates, (second row) instructions to
create objects and arrays, and to consult the array length, (third
row) read and write access to object fields, and, (fourth row) read
and write access to an array position.
%
%
As regards exceptions, they can be handled by treating them as
additional nodes and arcs in the control flow graph of the program.
In our framework, such flows are represented in the CLP-decompiled
program with explicit calls to the corresponding exception
handlers. 


For simplicity, the language does not include features of OO
imperative languages like bitwise operations, static fields, access
control (e.g., the use of {\sf public}, {\sf protected} and {\sf
  private} modifiers) and primitive types besides integers and
booleans.
Most of these features can be easily handled in this framework, as
shown by the implementation based on actual Java bytecode.

\secbeg
\subsection{Semantics of CLP-Decompiled Programs with
  Heap}\label{sec:semant-clp-decomp}
\secbeg

When considering a simple imperative language without heap-allocated
data structures, like in \cite{tdg-lopstr08}, CLP-decompiled programs
can be executed by using the standard execution mechanism of CLP. In
order to extend this approach to a realistic language with dynamic
memory, as our first contribution, we provide a suitable
representation for the heap and define the heap related
operations. Note that, in CLP-decompiled programs the heap is treated
as a black-box through its associated operations, therefore it is
always a variable. At run-time, the heap is represented as a list of
locations which are pairs made up of a unique reference and a cell,
which in turn can be an object or an array. An object contains its
type and its list of fields, each of them contains its signature and
data contents. An array contains its type, its length and the list of
its elements. Observe that arrays are stored in the heap together with
objects (as it happens e.g. in Java bytecode). 
Formally, the syntax of the heap at run-time is as follows. The
asterisks 
will be explained later:

\vspace{-.3cm}
\[ \small
\begin{array}{rllrll}
Heap &::=& \mbox{[] $|$ [{\it Loc$|$Heap}]} &  Cell &::=& \mbox{{\it object}({\it C$^*$,Fields$^*$}) $|$
                 {\it array}({\it T$^*$,Num$^*$,Args$^*$})}  \\
Loc &::=& \mbox{({\it Num$^*$,Cell}) }  & 
Fields &::=& \mbox{[] $|$ [{\it field}({\it FN,Data$^*$})$|${\it Fields$^*$}]} \\
\end{array}
\]
\vspace{-.2cm}

In the upper side of the figure, we present the
CLP-implementation of the operations to create heap-allocated data
structures (like \newobject\ and \newarray) and to read and modify
them (like \setfield, etc.), and, at the bottom appear some auxiliary
predicates.  To simplify the presentation some predicates are omitted,
namely: 
{\sf build\_object/2} resp. {\sf build\_array/3}, which create an
object, resp. an array term, {\sf new\_ref/1} which produces a fresh
numeric reference, and \textsf{\subclass/2} which implements the
transitive and reflexive \emph{subclass} relation on two classes.
\memberdet/2 resp. \replacedet/4 implements the usual deterministic
\emph{member}, resp. \emph{replace}, on lists, while \textsf{nth0/3}
resp. \textsf{replace\_nth0/3} implements the access to,
resp. replacement of, the i$^{th}$ element of a list using constraints
(multi-moded versions).

\begin{figure}[t]

\begin{center}
\small 
\setlength\fboxsep{1pt}
\fbox{
\begin{tabular*}{0.97\textwidth}{@{\extracolsep{\fill}}l}
\(\!
\begin{array}{@{}r@{~}c@{~}l}
  \mbox{\sf \newobject(H,C,Ref,H')}&\mbox{:-}& 
  \mbox{\sf build\_object(C,Ob), new\_ref(Ref), H' = [(Ref,Ob)\textbar H].}\\
  \mbox{\sf \newarray(H,T,L,Ref,H')}&\mbox{:-}& 
  \mbox{\sf  build\_array(T,L,Arr), new\_ref(Ref), H' = [(Ref,Arr)\textbar H].}\\[1ex]
  \mbox{\sf \type(H,Ref,T)}&\mbox{:-}& \mbox{\sf \getcell(H,Ref,Cell), Cell = object(T,\_).}\\
  \mbox{\sf \length(H,Ref,L)}&\mbox{:-}& \mbox{\sf \getcell(H,Ref,Cell), Cell = array(\_,L,\_).}\\[1ex]
  \mbox{\sf \getfield(H,Ref,FSig,V)}&\mbox{:-}& \mbox{\sf \getcell(H,Ref,Ob), 
            FSig = C:FN, Ob = object(T,Fields),}\\
  & & \mbox{\sf \subclass(T,C), \memberdet(field(FN,V),Fields).}\\
\mbox{\sf get\_array(H,Ref,I,V)}&\mbox{:-}&
\mbox{\sf \getcell(H,Ref,Arr), Arr = array(\_,\_,Xs), nth0(I,Xs,V).}\\[1ex]
\mbox{\sf \setfield(H,Ref,FSig,V,H')} & \mbox{:-}&
\mbox{\sf \getcell(H,Ref,Ob), FSig = C:FN, Ob = object(T,Fields), }\\
& &\mbox{\sf \subclass(T,C), \replacedet(Fields,field(FN,\_),field(FN,V),Fds'),} \\
& & \mbox{\sf \setcell(H,Ref,object(T,Fds'),H').}\\
\mbox{\sf set\_array(H,Ref,I,V,H')} &\mbox{:-}&
\mbox{\sf \getcell(H,Ref,Arr), Arr = array(T,L,Xs),}\\
& & \mbox{\sf replace\_nth0(Xs,I,V,Xs'), \setcell(H,Ref,array(T,L,Xs'),H').}\\[1ex]
\end{array}
\) \vspace{-.3cm}\\\hline
\(
\begin{array}{l}
\mbox{\sf \getcell([(Ref',Cell')\textbar \_],Ref,Cell) :- Ref ==
  Ref', !, Cell = Cell'.} \\
\mbox{\sf \getcell([\_\textbar RH],Ref,Ob) :- \getcell(RH,Ref,Ob).}\\[1ex]
\mbox{\sf \setcell([(Ref',\_)\textbar H],Ref,Cell,H') :- 
          Ref == Ref', !, H' = [(Ref,Cell)\textbar H].}\\
\mbox{\sf \setcell([(Ref',Cell')\textbar H'],Ref,Cell,H) :-
          H = [(Ref',Cell')\textbar H''],  
          \setcell(H',Ref,Cell,H'').}\\[1ex]
\end{array}
\) 
\end{tabular*}
}
\end{center}
\secbeg\secbeg
\caption{Heap operations for ground execution}
\label{fig:heap-ops}
\end{figure}

We now focus on the \emph{ground} execution of CLP-decompiled programs
in which we assume that all input parameters of the predicate to be
executed (i.e., $Args_{in}$ and $H_{in}$) are fully instantiated.  The
instantiations are provided as constraints in the \emph{input state}.
We assume familiarity with the basic notions of CLP. Very briefly, let
us recall that the operational semantics of a CLP program $P$ can be
defined in terms of \emph{derivations}, which are sequences of
reductions between \emph{states} $S_0 \rightarrow_P S_1 \rightarrow_P
... \rightarrow_P S_n$, also denoted $S_0\rightarrow^*_P S_n$, where a
{\em state} $\tuple{G\gd{}\theta}$ consists of a goal $G$ and a
constraint store $\theta$. If the derivation successfully terminates,
then $S_n=\tuple{\epsilon \gd{}\theta'}$ and $\theta'$ is called the
\emph{output state}.

\begin{definition}[ground execution]
  Let $M$ be a method, $m$ be the corresponding predicate from its
  associated CLP-decompiled program $P$, and $P'$ be the union of $P$
  and the clauses in Fig.~\ref{fig:heap-ops}. The \emph{ground
    execution} of $m$ with input $\theta$ is the derivation $S_0
  \rightarrow_{P'}^* S_n$, where $S_0 = \tuple{m(\mbox{\it
      Args$_{in}$,Args$_{out}$,H$_{in}$,H$_{out}$,ExFlag})
    \gd{}\theta}$ and $\theta$ initializes $Args_{in}$ and $H_{in}$ to
  be fully ground. If the derivation successfully terminates, then
  $S_n=\tuple{\epsilon \gd{}\theta'}$ and $\theta'$ is the output
  state ($\epsilon$ denotes the empty goal).
\end{definition}
Every CLP-decompilation must ensure that \clp programs capture the
same semantics of the original imperative ones.  This is to say that,
given a \emph{correct input state}, the CLP-execution yields an
\emph{equivalent} output state.
%
By \emph{correct} input state, we mean that all input arguments have
the correct types and that the heap has the required contents. For
instance, \(\theta = \{\mbox{\sf Args$_{in}$ = [r(1),null]} \wedge
\mbox{\sf H$_{in}$ = }\) {\sf
  [(1,object('SL',[field('SL':first,null)]))]\}} is a correct input
state for predicate \texttt{merge/5}, whereas $\theta = \{\mbox{\sf
  Args$_{in}$ = [r(1),r(2)]} \wedge \mbox{\sf H$_{in}$ = []}\}$ is not
correct since the heap does not include the required objects.
%

\begin{definition}[correct decompilation]\label{def:correct-decomp}
  Consider a method $M$ and a correct input state $I$.  Let $m$ be the
  CLP-decompiled predicate obtained from $M$ and $\theta$ be the input
  state equivalent to $I$.  If the CLP-decompilation is \emph{correct}
  then it must hold that, the execution in the OO language of $M$
  returns as output state $O$ if and only if the ground execution of
  $m$ with $\theta$ is deterministic and returns an output state
  $\theta'$ equivalent to $\sf O$.
\end{definition}
Correctness must be proven for the particular techniques used to carry
out the decompilation. In the interpretive approach, for a simpler
bytecode language without heap, \cite{mod-decomp-jist09} proves that
the execution of the decompiled programs produces the same output
state than the execution of the bytecode program in the CLP
interpreter. A full proof would require to prove that the CLP
interpreter is correct and complete w.r.t the corresponding imperative
language semantics. Since our approach is not tied to a particular
decompilation technique, in the rest of the paper, for the correctness
of our \tdg approach, we just require that decompiled programs are
correct as stated in Def.~\ref{def:correct-decomp}.
%

Finally, in the above definition, it can be observed that, since
CLP-decompiled programs originate from imperative bytecode, their
ground execution is deterministic. The aim of the next section is to
be able to execute CLP-decompiled programs symbolically with the input
arguments being free variables.

\secbeg

\secbeg

\section{Symbolic Execution of OO Imperative
  Programs}\label{sec:using-clp-test}


Interestingly, our CLP-decompiled programs can in principle be used,
not only to perform ground execution, but also \emph{symbolic
  execution} (\syex). Indeed, when the imperative language does not
use dynamic memory nor OO features, we can simply run the
CLP-decompiled programs by using the standard CLP execution mechanism
with all arguments being distinct free variables. For simple
imperative languages, this approach was first proposed by
\cite{Meudec01} and developed for a simple bytecode language in
\cite{tdg-lopstr08}. However, dealing with dynamic memory and OO
features entails further complications, as we show in this section.
%

\secbeg
\subsection{Handling Heap-Allocation in Symbolic Execution}\label{sec:symb-exec-with}
\secbeg

In principle, \syex starts with a fully unknown input state, including
a fully unknown heap. Thus, one has to provide some method which
builds a heap associated with a given path by using only the
constraints induced by the visited code. In the case of \tcg, it is
required that the ground execution with that heap (and the
corresponding input arguments) traverses exactly such path.
Existing approaches define novel specific operators to carry out this
task. For instance, \cite{CharreteurBG09} adds new constraint models
for the heap that extend the basic constraint-based approach without
heap. Similarly, \cite{Schrijvers-LOPSTR09} provides specific
constraints for heap-allocated lists, but needs to adjust the solver
to handle other data structures.
In our approach, thanks to the explicit representation of the heap,
we are able to provide a general solution for the \syex of programs
with arbitrary heap-allocated data structures.

The main point is that in a ground execution, the heap is totally
instantiated and, when we execute \getcell/3 (see
Fig.~\ref{fig:heap-ops}), the reference we are searching for must be a
number (not a variable) existing in the heap. In contrast, \syex deals
with partially unknown heaps. Our solution consists in generalizing
the definition of \getcell/3 by adding an additional clause (the first
one) as follows:
\vspace{-.1cm}
\[\small
\begin{array}{lll}
\mbox{\sf \getcell(H,Ref,Cell)} &\mbox{:-}& \mbox{\sf var(H), !, H = [(Ref,Cell)\textbar \_].}\\
\mbox{\sf \getcell([(Ref',Cell')\textbar \_],Ref,Cell)} &\mbox{:-}&
\mbox{\sf Ref == Ref', !, Cell = Cell'.}\\
\mbox{\sf \getcell([\_\textbar RH],Ref,Cell)}&\mbox{:-}& \mbox{\sf \getcell(RH,Ref,Cell).}\\
\end{array}
\] 

\noindent Intuitively, the heap during \syex contains two parts: the
\emph{known part}, with the cells that have been explicitly created
during \syex which appear at the beginning of the list, and the
\emph{unknown part}, which is a logic variable (tail of the list) in
which new data can be added. Observe the syntax of the heap in
Sect.~\ref{sec:semant-clp-decomp} where the *'s indicate where partial
information can occur in the heaps during \syex. Such syntax is hence
valid for all heaps appearing at \syex time. The definition of
\getcell/3 now distinguishes two situations when searching for a
reference: (i) It finds it in the known part (second clause). Note the
use of syntactic equality rather than unification since references at
\syex time can be variables or numbers.
(ii) Otherwise, it reaches the unknown part of the heap (a logic
variable), and it allocates the reference (in this case a variable)
there (first clause).

\begin{example}\label{ex:tcg}
  Let us use our \syex framework for the purpose of \tcg on our
  working example. As will be further explained, for this it is
  required to: (i) impose a termination criterion on \syex, and (ii)
  have a mechanism to produce actual values from the obtained path
  constraints. For (i) let us use \bck{k} with
  $K=2$. 
  Regarding (ii), we just rely on the \emph{labeling} mechanism of
  standard \emph{clpfd} domains, since we only get arithmetic path
  constraints. The rest of the constraints are handled as explained
  with standard unification through the defined heap operations.
  Table~\ref{table:example-testcases} depicts a graphical
  representation of the obtained set of test-cases.
%
  The table shows, for each test-case, an identifier, a graphical
  representation of its input and output, and the exception flag. Due
  to space limitations, we do not show the full input and output
  heaps, but instead we use the customary graphical representation for
  the linked lists of integers that they contain (see the example
  below to understand the correspondence).
  Let us focus on the first test-case. It corresponds to the following
  (simplified) sequence of reduction steps $\small\mbox{\tt merge
    $\!\!\rightarrow\!\!$ nullcheck$_1 \!\rightarrow\!$
    nullcheck$_2\!\!\rightarrow\!\!$ nullcheck$_3\!\!\rightarrow\!\!$
    if1$_1\!\!\rightarrow\!\!$ preloop $\!\!\rightarrow\!\!\!$ loop
    $\!\!\rightarrow\!\!\!$ loopcond1$_2\!\!\!\rightarrow$}$
  $\small\mbox{\tt $\rightarrow\!\!$ loopcond2$_1\!\!\rightarrow\!\!$
    if3$_1$}$. Its associated answer is \(\footnotesize{\theta =
    \{\mbox{\sf Args$_{in}$ = [r(Th),r(L)]} \wedge \mbox{\sf H$_{in}$
      =}}\) \footnotesize{\sf [(Th, object('SL',[field(first,A)])),
    (L,object('SL',[field(first,B)])),(A,object('SLNode',[field(data,1)])),
    (B, object('SLNode',[field(data,0),field(next,null)]))] $\wedge
    \ldots \}$}\normalsize ,
  indicating that merging a list with head ``{\sf 1}'' and any
  possible continuation (denoted ``\textsf{C}''), and a
  null-terminated list with head ``{\sf 0}'', produces an output list
  with head ``\textsf{0}'', followed by ``\textsf{1}'' and followed by
  the continuation ``\textsf{C}''.
%
  The last three test-cases show that, either if \texttt{l} is
  \textsf{null}, or the \texttt{first} field of any of the lists is
  \textsf{null}, the method throws an exception. This is indeed
  spotting a bug in the program (assuming it is not the intended
  behavior).
\end{example}

\begin{table}[t]
\setlength\fboxsep{0pt}
\setlength{\tabcolsep}{1.5mm}
\renewcommand{\arraystretch}{1}
\begin{center}

\footnotesize
\fbox{
\entrymodifiers={+[F-:<5pt>]}
\begin{tabular*}{0.97\textwidth}{@{\extracolsep{\fill}}ll@{~~~~}ll}
N & Input & Output & EF \\ \toprule

1 & \(\xymatrix@!C=0.1cm{
*{\sf this.first}\ar[r] & 1\ar[r] & *{~C}}
~~~~\xymatrix@!C=0.6pt{*{\sf l.first}\ar[r] & 0\ar[r] & *{\sf ~null}}\)
& \(\xymatrix@!C=0.1cm{
*{\sf this.first}\ar[r] & 0\ar[r] & 1\ar[r] & *{~C}}\)
& ok \\ \midrule

2 & \(\xymatrix@!C=0.1cm{
*{\sf this.first}\ar[r] & 1\ar[r] & *{~C}}\)
& \(\xymatrix@!C=0.1cm{
*{\sf this.first}\ar[r] & 0\ar[r] & 0\ar[r] & 1\ar[r] & *{~C}}\)
& ok \\
& \(\xymatrix@!C=0.6pt{
*{\sf l.first}\ar[r] & 0\ar[r] & 0\ar[r] & *{\sf ~null}}\)\\ \midrule

3 & \(\xymatrix@!C=0.1cm{
*{\sf this.first}\ar[r] & 1\ar[r] & *{\sf ~null}}\)
& \(\xymatrix@!C=0.1cm{
*{\sf this.first}\ar[r] & 0\ar[r] & 1\ar[r] & 1\ar[r] & *{~C}}\)
& ok \\
& \(\xymatrix@!C=0.6pt{
*{\sf l.first}\ar[r] & 0\ar[r] & 1\ar[r] & *{~C}}\)\\ \midrule

4 & \(\xymatrix@!C=0.1cm{
*{\sf this.first}\ar[r] & 0\ar[r] & *{\sf ~null}}
~~~~\xymatrix@!C=0.6pt{*{\sf l.first}\ar[r] & 0\ar[r] & *{~C}}\)
& \(\xymatrix@!C=0.1cm{
*{\sf this.first}\ar[r] & 0\ar[r] & 0\ar[r] & *{~C}}\)
& ok \\ \midrule

5 & \(\xymatrix@!C=0.1cm{
*{\sf this.first}\ar[r] & 0\ar[r] & 1\ar[r] & *{~C}}\)
& \(\xymatrix@!C=0.1cm{
*{\sf this.first}\ar[r] & 0\ar[r] & 0\ar[r] & 1\ar[r] & *{~C}}\)
& ok \\
& \(\xymatrix@!C=0.6pt{
*{\sf l.first}\ar[r] & 0\ar[r] & *{\sf ~null}}\)\\ \midrule

6 & \(\xymatrix@!C=0.1cm{
*{\sf this.first}\ar[r] & 0\ar[r] & 0\ar[r] & *{\sf ~null}}\)
& \(\xymatrix@!C=0.1cm{
*{\sf this.first}\ar[r] & 0\ar[r] & 0\ar[r] & 0\ar[r] & *{~C}}\)
& ok \\
& \(\xymatrix@!C=0.6pt{
*{\sf l.first}\ar[r] & 0\ar[r] & *{~C}}\)\\ \midrule

7 & \(\xymatrix@!C=0.2cm{
*{\sf this.first}\ar[r] & 0\ar[r] & *{~C} &
*{\sf ~~~~~~l.first = null}}\)
& -
& exc\\ \midrule

8 & \(\xymatrix@!C=.3cm{
*{\sf this.first}\ar[r] & *{\sf ~null} &
*{\sf l.first}\ar[r] & *{\sf ~C}}\)
& -
& exc \\ \midrule

9 & \(\xymatrix@!C=0.6pt{
*{\sf this.first}\ar[r] & *{\sf ~C} & 
*{\sf l~}\ar[r] & *{\sf ~null}}\)
& -
& exc \\

\end{tabular*}
\entrymodifiers={}
}
\end{center}
\secbeg\secbeg
\caption{Obtained test-cases for working example}
\label{table:example-testcases}
\end{table}


\secbeg
\secbeg
\secbeg
\secbeg
\subsection{Handling Pointer Aliasing in Symbolic Execution}
\secbeg

A challenge in \syex of realistic languages is to consider
pointer-\emph{aliasing} during the generation of heap-allocated data
structures, i.e., the fact that the same memory location can be
accessed through several references (called aliases). In the case of
\tcg, ignoring aliasing can lead to a loss of coverage.
%
Again, our solution
consists in further generalizing the definition of \getcell/3 by adding
an additional clause (the third one), thus illustrating again the
flexibility of our approach:

\vspace{-.3cm}
\[ \small
\begin{array}{lll}
\mbox{\sf \getcell(H,Ref,Cell)} &\mbox{:-}& \mbox{\sf var(H), !, H = [(Ref,Cell)\textbar \_].}\\
\mbox{\sf \getcell([(Ref',Cell')\textbar \_],Ref,Cell)} &\mbox{:-}&
\mbox{\sf Ref == Ref', !, Cell = Cell'.}\\
\mbox{\sf \getcell([(Ref',Cell')\textbar \_],Ref,Cell)} &\mbox{:-}&
\mbox{\sf var(Ref), var(Ref'), Ref = Ref', Cell = Cell'.}\\
\mbox{\sf \getcell([\_\textbar RH],Ref,Cell)}&\mbox{:-}& \mbox{\sf \getcell(RH,Ref,Cell).}\\
\end{array}
\]

\noindent Essentially, two cases are distinguished: (a) The
reference we are searching for is a number, in that case it must
exist in the heap and the 2nd clause will eventually succeed.  (b)
If {\sf Ref} is a variable: (b.1) {\sf Ref} exists in the heap, and
the 2nd clause eventually succeeds.  Here, {\sf Ref}
must have been already processed (and possible aliases for it might
have been created. (b.2) The interesting case is when {\sf Ref} is a
free variable which was not in the heap.  In this case, the 2nd
clause will never succeed and the 3rd one will unify {\sf Ref} with
all matching references in the heap.

\begin{table}[t]
\setlength\fboxsep{0pt}
\setlength{\tabcolsep}{1.5mm}
\renewcommand{\arraystretch}{1}
\begin{center}

\footnotesize
\fbox{
\entrymodifiers={+[F-:<5pt>]}
\begin{tabular*}{0.95\textwidth}{@{\extracolsep{\fill}}llll}

N & Input & Output & EF \\ \toprule

10 & \(\xymatrix@!C=0.1cm@R=0.1cm{
*{\sf this.first}\ar[r] & 0\ar[r] & *{\sf ~null} &
*{~} & *{\sf l = this}}\) & 
\(\xymatrix@!C=0.2cm{
*{\sf this.first}\ar[r] & 0\ar@(r,d)[]}\) & ok \\ \midrule

11 & \(\xymatrix@!C=0.1cm@R=0.1cm{
*{\sf this.first}\ar[r] & 0\ar[r] & 0\ar[r] & *{\sf ~null} &
*{~} & *{\sf l = this}}\) & 
\(\xymatrix@!C=0.2cm{
*{\sf this.first}\ar[r] & 0\ar[r] & 0\ar@/^/[l] }\) & ok \\ \midrule

12 & \(\xymatrix@!C=0.2cm@R=0.1cm{
*{\sf this.first}\ar[r] & *{\sf ~null} &
*{\sf ~~~~l = this}}\) & - & exc \\ \midrule

13 & \(\xymatrix@!C=0.1cm@R=0.1cm{
*{\sf this.first}\ar[r] & 0\ar[r] & *{\sf ~null}\\
*{\sf l.first}\ar[ur]}\) & 
\(\xymatrix@!C=0.2cm{
*{\sf this.first}\ar[r] & 0\ar@(r,d)[]}\) & ok \\ \midrule

14 & \(\xymatrix@!C=0.1cm@R=0.1cm{
*{\sf this.first}\ar[r] & 0\ar[r] & 0\ar[r] & *{\sf ~null}\\
*{\sf l.first}\ar[ur]}\) & 
\(\xymatrix@!C=0.2cm{
*{\sf this.first}\ar[r] & 0\ar[r] & 0\ar@/^/[l] }\) & ok \\ \midrule

15 & \(\xymatrix@!C=0.1cm@R=0.05cm{
*{\sf this.first}\ar[r] & 1\ar[r] & *{\sf ~null}\\
*{\sf l.first}\ar[r] & 0\ar@(r,dr)@/_1.2pc/[u]}\) & 
\(\xymatrix@!C=0.2cm{
*{\sf this.first}\ar[r] & 0\ar[r] & 1\ar@(r,d)[]}\) & ok \\ \midrule

16 & \(\xymatrix@!C=0.1cm@R=0.05cm{
*{\sf this.first}\ar[r] & 0\ar@(r,ur)@/^1.2pc/[d] \\
*{\sf l.first}\ar[r] & 0\ar[r] & *{\sf ~null}}\) & 
\(\xymatrix@!C=0.2cm{
*{\sf this.first}\ar[r] & 0\ar[r] & 0\ar@(r,d)[]}\) & ok \\ 

\end{tabular*}
\entrymodifiers={}
}
\end{center}
\secbeg\secbeg
\caption{Additional test-cases when considering pointer-aliasing}
\label{table:example-testcases-aliasing}
\end{table}

\begin{example}\label{ex:aliasing}
  Let us consider again the \tcg for our working example as in
  Ex.~\ref{ex:tcg}. Table~\ref{table:example-testcases-aliasing} shows
  seven additional test-cases obtained using the new definition of
  \getcell/3. Test-cases 10-12 represent executions in which the two
  lists to be merged are aliases. The remaining test-cases show other
  shapes of lists with aliasing among their nodes. 
  In most cases, the result is a cyclic list. This clearly reveals a
  dangerous behavior of the method which should be controlled by the
  programmer.  Altogether, our set of test-cases provides full
  coverage w.r.t.\ the shape of data structures.
\end{example}

\subsection{Inheritance and Virtual Invocations in Symbolic
  Execution} \label{oo-features}
\secbeg

Inheritance and virtual method invocations pose further challenges in
\syex of realistic OO programming languages. From the side of data
structure shape coverage, we should create aliasing among objects that
possibly have different class types but, due to their inheritance
relation, might be aliased at runtime.  From the side of path
coverage, virtual invocations pose further complications when the
object on which the virtual invocation is performed has not been
created during \syex, but is rather accessed from the input
arguments. In this case, only the declaration type of the object is
known. To achieve path coverage, all implementations of the method
that might be invoked at runtime (but not more), should be exercised.
Interestingly, our solution solves these issues for free.
Let us consider a scenario where we have three classes $A$, $B$ and
$C$, such that $C$ is a subclass of $B$, and $B$ a subclass of $A$;
and the following method \( \mbox{\tt m(A a, B
  b)\{a.f;~b.g;~a.p();\}}\). Let us also assume that both $B$ and $C$
redefine method $p$. The corresponding \clp-decompiled code contains
two calls to \getfield/4, resp. with \texttt{'A':f} and
\texttt{'B':g}. During \syex, the first one will call
\textsf{\subclass(X,'A')}, which produces three alternatives
(\textsf{X='A', X='B' and X='C'}). The second call to \getfield\ will
then succeed with cases \textsf{X='B'} and \textsf{X='C'}, but fail
with \textsf{X='A'}. Thus, the case where $a$ and $b$ are aliased is
properly handled, and the calls $B.p()$ and $C.p()$ (and not $A.p()$)
will be exercised.

\begin{definition}[symbolic execution]
  Let $M$ be a method, $m$ be the corresponding predicate from its
  associated CLP-decompiled program $P$, and $P'$ be the union of $P$
  and the clauses in Fig.~\ref{fig:heap-ops} with the described
  extensions. The \emph{symbolic execution} of $m$ is the derivation
  tree with root $S_0 = \tuple{m(\mbox{\it
      Args$_{in}$,Args$_{out}$,H$_{in}$,H$_{out}$,ExFlag})
    \gd{}\theta}$ and $\theta = \{\}$ obtained using $P'$.
\end{definition}
%
%
The following theorem establishes the correctness of our symbolic
execution mechanism. Intuitively, it says that each successful
derivation in the symbolic execution produces an output state which is
correct, i.e., for any ground instantiation of such derivation we
obtain an output state which is an instantiation of the one obtained
in the symbolic execution. For simplicity, throughout the paper, we
have included in an output state $\theta$ two ingredients: the
computed answer substitution $\sigma$ and the actual constraints
$\gamma$. Given a constraint store $\theta$, we say that $\sigma'$ is
an \emph{instantiation} of $\theta$ if $\sigma'\leq \sigma$ and
$\gamma\sigma'$ is satisfiable. Also, we say that an output state
$\theta'$ is an instantiation of $\theta$, written $\theta' \leq
\theta$, when both the corresponding stores and the substitutions hold
the $\leq$ relation.


\begin{theorem}[correctness] 
  Consider a successful derivation of the form: $S_0 \rightarrow S_1
  \rightarrow ... \rightarrow \tuple{\epsilon \gd{}\theta}$ which is a
  branch of the tree with root $S_0 = \tuple{m(\mbox{\it
      Args$_{in}$,Args$_{out}$,H$_{in}$,H$_{out}$,ExFlag}) \gd{}\{\}}$
  obtained in the \emph{symbolic execution} of $m$. Then, for any
  instantiation $\sigma'$ of $\theta$ which initializes $Args_{in}$ and
  $H_{in}$ to be fully ground, it holds that the ground execution of
  $S'_0 = \tuple{m(\mbox{\it
      Args$_{in}$,Args$_{out}$,H$_{in}$,H$_{out}$,ExFlag})\sigma'
    \gd{}\{\}}$ results in  $\tuple{\epsilon \gd{}\theta'}$
  with $\theta' \leq \theta$.





 \end{theorem}

\secbeg

\secbeg

\section{(Conditional) \tcg of OO Imperative Programs}\label{sec:impl-stand-cover}

An important problem with \syex, regardless of whether it is performed
using CLP or a dedicated execution engine, is that the execution tree
to be traversed is in general infinite. 
In the context of \tcg, it is therefore essential to establish a
\emph{termination criterion},
which guarantees that the number of paths traversed remains finite,
while at the same time an interesting set of test-cases is generated.
In addition to this, some approaches perform \emph{conditional} \tcg
in which, besides selecting a criterion, the user establishes a
precondition which further prunes the evaluation tree. In the
remaining of this section, we describe how these issues are handled in
our approach.

\subsection{Implementing Coverage Criteria by means of Unfolding
  Strategies}\label{sec:cov_crit}

A large series of \emph{coverage criteria} (\ccs) have been developed
over the years which aim at guaranteeing that the program is exercised
on interesting control and/or data flows. Applying the coverage
criteria on the CLP-decompiled program should achieve the desired
coverage on the original bytecode. 
%

Implementing a \cc in our approach consists in building a finite
(possibly unfinished) evaluation tree by using a non-standard
evaluation strategy.  In \cite{tdg-lopstr08}, we observed that this is
exactly the problem that \emph{unfolding rules} used in partial
evaluators of (C)LP solve, and we proposed \bck{k}, a new \cc for
bytecode which was implemented with the corresponding unfolding
rule. In this section, we go further and show that the most common
\ccs can be integrated in our system using unfolding rules. The
following predicate defines a generic unfolding rule for depth-first
evaluation strategies which is parametric w.r.t. the \cc:
\[\small\!\!\!
\begin{array}{l}
\mbox{\tt unfold(Root,Goal,CCAuxDS,CCParam) :-}\\
\mbox{\tt {\tiny (1)}~~~select(Goal,G$_{\tt left}$,A,G$_{\tt right}$), !,}\\
\mbox{\tt {\tiny (2)}~~~(internal(A) -> match(A,Bs) ; (call(A), Bs = []),}\\
\mbox{\tt {\tiny (3)}~~~update\_ccaux(CCAuxDS,A,CCAuxDS'),}\\
\mbox{\tt {\tiny (4)}~~~append([G$_{\tt left}$,Bs,G$_{\tt right}$],Goal'),}\\
\mbox{\tt {\tiny (5)}~~~(terminates(A,CCAuxDS',CCParam) -> add\_resultant(Root,Goal') }\\
\mbox{\tt {\tiny (6)}~~~~~~~~~~~~~~~~~~~~~~~~~~~~~~~~~~~~; 
          unfold(Root,Goal',CCAuxDS',CCParam)).}\\
\mbox{\tt unfold(Root,Goal,\_,\_) :- add\_resultant(Root,Goal).}\\
\end{array}
\]

\noindent The main operation dependent on the \cc is
\texttt{terminates/3}, which indicates when the derivation must be
stopped. For this aim, it uses an input set of parameters
\texttt{CCParam} and an auxiliary data-structure \texttt{CCAuxDS}.
Intuitively, given a goal \texttt{Goal}, an initial \texttt{CCAuxDS}
and \texttt{CCParams}, \texttt{unfold/4} performs unfolding steps
until either \texttt{select/4} fails, because there are no atoms to be
reduced in the goal, or \texttt{terminates/3} succeeds. In both cases,
the corresponding \emph{resultant} is stored, which can then be used
to generate a test-case (or a rule in the \emph{test-case generator}
\cite{tdg-lopstr08}). The \texttt{Root} argument carries along the
root atom of \syex. An unfolding step consists in the following: (1)
select the atom to be reduced, which splits the goal into the selected
atom \texttt{A} and the sub-goals to its left \texttt{G$_{\tt left}$}
and right \texttt{G$_{\tt right}$}; (2) match the atom with the head
of a clause in the program, or call it in case it is a builtin or
constraint; (3) update \texttt{CCAuxDS}; (4) compose the new goal; and
(5) if the \cc stops the derivation (i.e. \texttt{terminates/3}
succeeds) then store the resultant, otherwise (6) continue unfolding.

In order to instantiate this generic unfolding rule with a specific
\cc, one has to provide the corresponding auxiliary data-structure and
parameters, as well as suitable implementations for
\texttt{update\_ccaux/3} and \texttt{terminates/3}. Additionally,
\texttt{match/2} and \texttt{select/4} allows resp. tuning the order
of generation of the evaluation tree, and extending the functionality
of \tcg by allowing \emph{non-leftmost} unfolding
steps~\cite{nonleftmost-lopstr05}, as will be further discussed. Note
that, in order to guarantee that we get correct results in presence of
non-leftmost unfoldings, predicates which are ``jumped over'' must be
\emph{pure} (see \cite{nonleftmost-lopstr05} for more details).
E.g., for \bck{k}, \texttt{CCParam} is just the $K$ and
\texttt{CCAuxDS} is the \emph{ancestor stack} (see
\cite{tdg-lopstr08}).
%


\subsection{Including Preconditions during \tcg}\label{sec:preconds}

In practice, it is also essential to prune horizontally the evaluation
tree in order to limit the number of test-cases obtained without
sacrificing interesting paths. The information used to perform this
task is usually provided by the user by means of preconditions on the
inputs, formulated using a set of pre-defined properties. These
properties can range from simple arithmetic constraints, to more
complex properties like \emph{sharing} or \emph{cyclicity} of
data-structures.
We consider two levels of properties. The first-level comprises
properties which can be executed beforehand thus being carried along
by the CLP engine, like equality and disequality constraints,
arithmetic constraints, etc. E.g., let us re-consider
Ex.~\ref{ex:aliasing}. We can specify the precondition that the lists
are not aliased simply by providing these literals at the beginning of
the goal ``$\mbox{\tt Args$_{in}$ = [r(Th),L], member(L,[null,r(L')]),
  Th \#\textbackslash= L'}$''.

The second level comprises properties that require a certain level of
instantiation on inputs in order to be executed. 
Depending on the property, \texttt{unfold/4} can either: perform
non-leftmost unfoldings until having the required instantiation, or
incrementally check the property as the corresponding structure is
being generated, or just delay the property check until the end of the
derivation. Interestingly, the different behaviors can be achieved
providing suitable implementations of \texttt{select/2}. Let us
re-consider again Ex.~\ref{ex:aliasing}. We can specify the
precondition that the lists do not share by providing this in the goal
``$\mbox{\tt Args$_{in}$ = [Th,L], noshare(Th,L)}$'', where predicate
\texttt{noshare/2} checks that the data transitively referenced from
\texttt{Th} do not share with that from \texttt{L}.

\secbeg

\secbeg

\section{Experimental Evaluation}

\begin{table}[t]
\centering \scriptsize
\renewcommand{\arraystretch}{.4}
\setlength{\tabcolsep}{.75mm}
\begin{tabular*}{1.00\textwidth}{@{\extracolsep{\fill}}lcccccccccccccc}\toprule
\textbf{Bench} & \textbf{Es} & \textbf{Cs} & \textbf{Ms} &
\textbf{Is} & \textbf{T$_{dec}$} & 
\textbf{T$_{tcg}^{d50}$} & \textbf{N$^{d50}$} & \textbf{C$^{d50}$} & 
\textbf{T$_{tcg}^{d200}$} & \textbf{N$^{d200}$} & \textbf{C$^{d200}$} & 
\textbf{T$_{tcg}^{bk2}$} & \textbf{N$^{bk2}$} & \textbf{C$^{bk2}$} \\\toprule

Trityp & 1 & 1 & 1 & 98 & 38   & 22 & 14 & 100\% & 20 & 14 & 100\% & 22 & 14 & 100\% 
\\ \midrule
Josephus & 1 & 1 & 3 & 61 & 34   & 6 & 1 & 56\% & 366 & 45 & 100\% & 8 & 3 & 100\% 
\\ \midrule
DoublyLinkedList & 13 & 2 & 20 & 253 & 157   & 85 & 31 & 37\% & 594 & 178 & 100\% & 369 & 116 & 100\% 
\\ \midrule
RedBlackTree & 10 & 2 & 10 & 485 & 365   & 60 & 57 & 30\% & 2432 & 539 & 96\% & 10010 & 638 & 99\% 
\\ \toprule
NodeStack & 6 & 3 & 12 & 94 & 51   & 14 & 9 & 100\% & 8 & 9 & 100\% & 8 & 9 & 100\% 
\\ \midrule
ArrayStack & 7 & 3 & 11 & 103 & 58   & 16 & 15 & 100\% & 16 & 15 & 100\% & 16 & 15 & 100\% 
\\ \midrule
NodeQueue & 6 & 3 & 15 & 133 & 73   & 18 & 14 & 100\% & 13 & 15 & 100\% & 19 & 15 & 100\% 
\\ \midrule
NodeDeque & 9 & 3 & 19 & 223 & 150   & 32 & 23 & 67\% & 38 & 28 & 100\% & 34 & 28 & 100\% 
\\ \midrule
NodeList & 19 & 9 & 33 & 449 & 383   & 152 & 77 & 73\% & 182 & 91 & 91\% & 184 & 91 & 91\% 
\\ \midrule
SortedListPriorityQ\!\! & 11 & 14 & 40 & 491 & 442   & 62 & 33 & 29\% & 190 & 79 & 77\% & 512 & 164 & 91\% 
\\ \midrule
Sort & 4 & 9 & 30 & 735 & 661   & 26 & 12 & 12\% & 328 & 43 & 44\% & 400 & 55 & 72\% 
\\ \toprule
\end{tabular*}
\secbeg
\caption{Experimental results}
\label{tab:pet}
\end{table}

We have implemented and integrated the presented techniques in the
\pet tool~\cite{tdg-pepm10}, 
which is available for download and for online use through its web
interface at \texttt{http://costa.ls.fi.upm.es/pet}.
We now present some experiments which aim at illustrating the
applicability of our approach to \tcg of realistic OO programs.
We use two sets of benchmarks. The first group (first four benchmarks)
comprises a set of classical programs used to evaluate testing tools
taken from~\cite{jaut}. The second one (last seven) is a selection
from the \textsf{net.datastructures}
library~\cite{net-datastructures3}, a well-known library of algorithms
and data-structures for Java.
Table~\ref{tab:pet} shows the times taken by the different phases
performed by \pet as well as the number of test-cases generated and
the code coverage achieved for different \ccs, \bck{k} and \dpk{k}
(which simply limits the number of derivation steps). All times are in
milliseconds, and were obtained as the arithmetic mean of five runs on
an Intel Core 2 Quad Q9300 at 2.5GHz with 1.95GB of RAM, running Linux
2.6.26 (Debian lenny). For each benchmark we show: the number of
methods for which we have generated test-cases (\textbf{Es}); the
number of reachable classes, methods and Java bytecode instructions
(\textbf{Cs}, \textbf{Ms} and \textbf{Is}) (not considering Java
libraries); the time taken by \pet to decompile the bytecode to CLP
(\textbf{T$_{dec}$}); the time of the \tcg, total number of test-cases
and \emph{code coverage} for \dpk{50} (\textbf{T$_{tcg}^{d50}$},
\textbf{N$^{d50}$} and \textbf{C$^{d50}$}); for \dpk{200}
(\textbf{T$_{tcg}^{d200}$}, \textbf{N$^{d200}$} and
\textbf{C$^{d200}$}) and for \bck{2} (\textbf{T$_{tcg}^{bk2}$},
\textbf{N$^{bk2}$} and \textbf{C$^{bk2}$}).

The code coverage measures, given a method, the percentage of bytecode
instructions which are exercised by the obtained test-cases, among all
reachable instructions (including all transitively called
methods). This is usually the main measure considered in \tcg to
reason about the effectiveness of \ccs. We observe that \bck{2}
achieves a very high degree of coverage ($\simeq 100\%$ for the first
$8$ benchmarks) thus demonstrating its effectiveness in practice. There
are however cases where \bck{2} is not able to achieve $100\%$
coverage. There are different reasons for this: (i) In some cases, $K
= 2$ is not sufficient to reach some parts of the code. This is the
case of most methods in class \textsf{Sort}. Indeed, \bck{3} achieves
$100\%$ of code coverage for this class. (ii) Sometimes there are
parts of the code which are simply unreachable at execution time
(\emph{dead code}). This is frequent in very generic OO programs, as
it is the case of some methods reachable from \textsf{NodeList} and
\textsf{SortedListPriorityQ}.
 
The results obtained for \dpk{k} show that its effectiveness highly
depends on the chosen $k$, and this in turn depends on the particular
program. This results in an unsatisfactory \cc in practice. E.g.,
\dpk{50} for \textsf{Josephus} obtains $1$ test-case in $6$ ms, which
exercises only the $56\%$ of the code. However \dpk{200} achieves
$100\%$ coverage, but at the cost of spending much more time ($366$
ms), thus obtaining many more test-cases (45). Observe that \bck{2}
can achieve $100\%$ coverage with $3$ test-cases in only $8$
ms. 

Overall, from the first group of benchmarks we conclude that \pet can
compete and even outperform related
tools~\cite{jaut,DBLP:conf/tap/TillmannH08}. The second group
demonstrates the effectiveness of \pet with realistic OO programs
making extensive use of inheritance and virtual invocations.
A careful look at the most complex methods suggests that a more
restrictive \cc should be used to further prune the \syex tree when
considering more complex programs. E.g. \pet obtains
$276$ (in $880$ ms) for \textsf{RedBlackTree.fixAfterInsertion}. We
conclude also that the use of preconditions, as explained in
Sect.~\ref{sec:preconds}, (in principle provided by the user) will be
crucial in order to obtain manageable test-suites for more complex
programs. 
%

\secbeg

\secbeg

\secbeg
\secbeg

\section{Related Work and Conclusions}\label{sec:conclusions}
\secbeg

In the fields of program verification, static analysis and static
checking, transformational approaches are widely used
\cite{989788,DBLP:conf/tacas/VaziriJ03}. The common technique is to
translate an imperative program into an equivalent intermediate
representation on which the verification, analysis or checking is
performed. The work of \cite{989788} is similar to ours in the
translation of the imperative program into a constraint logic
one. However, the goal here is to perform bounded software model
checking rather than \tdg and it is not concerned with our problems of
ensuring coverage of the shape of data structures. Also, there are no
extensions to consider OO features like in our work. In the case of
\cite{DBLP:conf/tacas/VaziriJ03}, the imperative program is translated
into a propositional formula and SAT solving is used to find a
solution. Again, coverage of shape of data structures is not studied
here, which makes it fundamentally different from ours.

Much attention has been devoted to the use of constraint solving in
the automation of software testing since the seminal work
of~\cite{demillo-data-gen-91}. For the particular case of Java
bytecode, \cite{MullerLK04} develops a symbolic Java virtual machine
which integrates constraint solvers and a backtracking mechanism, as
without knowledge about the input data, the execution engine might
need to execute more than one path. In other approaches the problem is
tackled by transforming the program into corresponding constraints, on
which the testing process is then carried out by applying constraint
solving techniques. Recent progress has been done in this direction
towards handling heap-allocated data structures
\cite{271790,CharreteurBG09,Schrijvers-LOPSTR09}. An important
advantage of our approach is that, since the source program is
transformed into another (constraint logic) program --and not into
constraints only-- on which the symbolic execution is performed, we
can easily track the relation between the test-cases and the source
program. Keeping this relation is important for at least two reasons:
$(1)$ in order to model new coverage criteria on the source program by
using particular evaluation techniques on the CLP counterpart, and
$(2)$ to relate the generated test-cases to paths in the source
program to spot errors, etc. This relation is less clear in pure
constraint-based approaches (see discussion in
\cite{Schrijvers-LOPSTR09}). Some approaches are focused on improving
the efficiency of \tdg for dynamic pointer data
\cite{1307462,787004}. The basic idea is to separate the process of
generating the shape of the data structure to the one of generating
values for the fields of data. Our approach is similar to them in that
both process are also separated and, although actual experimentation
is needed, we believe that a similar efficiency will be achieved.

As another important point, while numeric data can be natively
supported by constraint solvers, when extending the constraint-based
approach to handle heap-allocated data structures, one has to define
new constraint models based on operators that model the heap
\cite{CharreteurBG09}. In \cite{Schrijvers-LOPSTR09}, these
constraints operators are implemented in CHR.
In these approaches, one needs to adjust the solver to the particular
data structures considered in the language. For instance,
\cite{Schrijvers-LOPSTR09} provides support for lists and sketches how
to extend it to handle trees by adding new operators.
Instead, we have provided a general solution to generate arbitrary
data structures by means of objects.
%

\secbeg

\subsection*{Acknowledgements}
We gratefully thank Samir Genaim and the anonymous referees for many
useful comments and suggestions that greatly helped improve this
article.  This work was funded in part by the Information \&
Communication Technologies program of the European Commission, Future
and Emerging Technologies (FET), under the ICT-231620 {\em HATS}
project, by the Spanish Ministry of Science and Innovation (MICINN)
under the TIN-2008-05624 {\em DOVES} project, the TIN2008-04473-E
(Acci\'on Especial) project, the HI2008-0153 (Acci\'on Integrada)
project, the UCM-BSCH-GR58/08-910502 Research Group and by the Madrid
Regional Government under the S2009TIC-1465 \emph{PROMETIDOS} project.

\bibliographystyle{acmtrans}


\end{document}